\documentclass[10pt,aps,twocolumn,prc,superscriptaddress,noshowpacs,nofootinbib,noshowkeys,floatfix]{revtex4}
\usepackage[dvips]{graphics,graphicx}
\usepackage[colorlinks=true,linktocpage=true,linkcolor=blue,citecolor=blue]{hyperref}
\usepackage[usenames,dvipsnames]{color}
\usepackage{amsmath, amssymb}
\usepackage{multirow}
\usepackage{longtable}
\usepackage{bm}
\usepackage{color}
\usepackage[normalem]{ulem}  % \sout{old text} for strikeout

\renewcommand\sout{\bgroup \color{blue} \ULdepth=-.5ex \ULset}
\begin{document}

\preprint{}

\title{Anisotropic escape mechanism and elliptic flow of bottomonia}

\author{Partha Pratim Bhaduri}
\affiliation{Variable Energy Cyclotron Centre, HBNI, 1/AF Bidhan Nagar, Kolkata 700 064, India}
\author{Nicolas Borghini}
\affiliation{Fakult\"at f\"ur Physik, Universit\"at Bielefeld, Postfach 100131, D-33501 Bielefeld, Germany}
\author{Amaresh Jaiswal}
\affiliation{School of Physical Sciences, National Institute of Science Education and Research, HBNI, Jatni-752050, India}
\author{Michael Strickland}
\affiliation{Department of Physics, Kent State University, Kent, OH 44242 United States}

\date{\today}

\begin{abstract}

We study the role of anisotropic escape in generating the elliptic flow of bottomonia produced in ultrarelativistic heavy-ion collisions. We implement temperature-dependent decay widths for the various bottomonium states, to calculate their survival probability when traversing through the anisotropic hot medium formed in non-central collisions. We employ the recently developed 3+1d quasiparticle anisotropic hydrodynamic simulation to model the space-time evolution of the quark-gluon plasma. We provide a quantitative prediction for transverse momentum dependence of bottomonium elliptic flow and nuclear modification factor for Pb$\,+\,$Pb collisions in $\sqrt{s_{\rm NN}}=2.76$~TeV at the Large Hadron Collider.

%We study the role of anisotropic escape in generating elliptic flow of bottomonium produced in ultrarelativistic heavy-ion collisions. We use the Glauber model to generate the initial energy density distribution in the plane transverse to the beam axis. We use previously determined temperature-dependent decay widths for bottomonium states which have been successful in described bottomonium suppression and calculate their survival probability when traversing through the hot and dense anisotropic medium formed in non-central collisions. We consider longitudinal Bjorken flow and mimic the transverse expansion of the fireball in an effective way. We calculate the transverse momentum dependence of elliptic flow generated from the anisotropic escape mechanism for mid-central Pb$\,+\,$Pb collisions in $\sqrt{s_{\rm NN}}=2.76$~TeV at the Large Hadron Collider. We provide a quantitative prediction for bottomonium elliptic flow generated from anisotropic escape mechanism.
\end{abstract}


% 25.75.-q Relativistic heavy-ion collisions
% 24.10.Nz Hydrodynamic models
% 47.75.+f Relativistic fluid dynamics

\maketitle

\section{Introduction}

Heavy quarks, such as charm ($c$) and bottom ($b$) quarks, and their quarkonium bound states ($c\bar{c}$ and $b\bar{b}$) are very useful internal probes of the hot and dense medium created in collisions of heavy nuclei at high energies~\cite{Andronic:2015wma}. Heavy quarkonia created in high-energy collisions at the BNL Relativistic Heavy Ion Collider (RHIC) and the CERN Large Hadron Collider (LHC) have been found to be appreciably affected by the medium. This leads to distinctive features in their observed final yields, which encode information about the thermodynamic and transport properties of the medium. Therefore understanding the dynamics of heavy quarks and quarkonia in a deconfined medium is of great interest for the heavy-ion physics community \cite{Matsui:1986dk, Karsch:1987pv, Brambilla:2004wf, vanHees:2004gq, vanHees:2005wb, Rapp:2008tf, Kluberg:2009wc, Andronic:2015wma}.

In the classical picture~\cite{Matsui:1986dk, Karsch:1987pv}, heavy quarkonia embedded in a static, equilibrated quark-gluon plasma (QGP) may survive at temperatures above the QGP crossover temperature due to their large binding energies. However, if the QGP energy density becomes sufficiently high, the resulting Debye screening of the quark-antiquark potential eventually leads to the dissociation of charmonia and bottomonia~\cite{Matsui:1986dk, Karsch:1987pv}. In this classical picture, the bound states with the largest binding energy, respectively the $J/\psi$ and the $\Upsilon$, have the highest dissociation temperatures. 

In recent years, this simple picture has been challenged by a number of findings. Based on first-principles finite-temperature quantum chromodynamics (QCD) calculations, it was shown that the in-medium quark-antiquark potential contains an imaginary part which is associated with the in-medium quarkonium breakup rate~\cite{Laine:2006ns, Dumitru:2007hy}. This results in significant thermal widths for quarkonia, at variance with the older assumption of uniquely defined binding energies, and leads to the suppression of quarkonia at temperatures at which they would survive in the traditional scenario~\cite{Strickland:2011mw,Strickland:2011aa,Krouppa:2015yoa, Krouppa:2016jcl, Krouppa:2017jlg}.

Another extension to the standard idea is the introduction of dynamics.  In the dynamical picture, quarkonium states are both dissociated and \mbox{(re-)associated} with the rate for each process depending on the open heavy flavor density and temperature of the system~\cite{Grandchamp:2003uw, Emerick:2011xu}. In particular, in an evolving QGP with decreasing temperature, a quark and an antiquark pair may combine into a stable bound state which was until then unstable~\cite{Thews:2000rj, BraunMunzinger:2000px}. For the fireball created in heavy-ion collisions at RHIC or LHC, recombination seems to be marginal for bottomonia~\cite{Du:2017qkv,Krouppa:2018lkt} while playing a more important role in the observed yields of charmonia \cite{Zhao:2010nk}. In this paper we will investigate bottomonia, whose dynamical evolution is not significantly affected by regeneration, due to the fact that $b\bar{b}$ pairs are less abundantly produced by initial hard scatterings than $c\bar{c}$ pairs.

An additional important aspect of the dynamics is that the dissociation process is not instantaneous, but depends on how long the quark-antiquark pairs experience a high medium energy density. This is irrelevant in a static infinite QGP, but becomes important in an expanding finite-sized fireball. In that case, and for pairs in motion relative to the QGP, this translates into a dependence on the in-medium path length of the pairs. Thus, the bound states created in a high energy heavy-ion collision that survive the ensuing immersion in the fireball are those that quickly reach a region of low enough temperature. This path-length dependence of the bound-state survival probability is naturally described within an ``escape mechanism'' scenario~\cite{Borghini:2010hy, He:2015hfa, Romatschke:2015dha, Jaiswal:2017dxp}. Since a generic heavy-ion collision gives rises to a spatially anisotropic overlap in the plane transverse to the beam axis, the path-length dependence of states propagating through the generated QGP results in an anisotropic emission pattern of the quarkonia, measured in momentum space, as was first predicted for $J/\psi$~\cite{Wang:2002ck}. For bottomonia, the anisotropic escape probability should constitute the major ingredient to the observed momentum anisotropy, quantified as usual in terms of Fourier harmonics $v_n$.

In this paper, we provide a quantitative prediction for bottomonium elliptic flow $v_2$ generated by the anisotropic escape mechanism; for a qualitative discussion, see Ref.~\cite{Das:2018xel}. In Sec.~\ref{s:model_fireball} we give the details of the model used to simulate the hydrodynamic evolution of the fireball created in Pb$\,+\,$Pb collisions at $\sqrt{s_{\rm NN}}=2.76$~TeV at the LHC. For the bottomonium states, we implement temperature-dependent decay widths and calculate their resulting survival probability in the hot and dense anisotropic medium in Sec.~\ref{s:model_survival}. Including the feed down contribution to the bottomonium ground state from higher excited states, we find that the elliptic flow of the $\Upsilon(1S)$ is at the level of a few percent and investigate its dependence on the parameters governing the medium expansion in Sec.~\ref{s:results}. We also study the dependence of the bottomonium elliptic flow on transverse momentum and centrality.

%%%%%%%%%%%%%%%%%%%%%%%%%%%%%%%%%%%%%%%%%%%%%%%%%%%%%%%%%%%%%%%%%%%%%

\section{Hydrodynamical model}
\label{s:model_fireball}

For the spatiotemporal evolution of the quark-gluon plasma we make use of the recently developed 3+1d quasiparticle anisotropic hydrodynamics (QaHydro) framework \cite{Alqahtani:2015qja}.  This framework has been successfully used to describe identified-particle spectra, charged-particle multiplicity vs pseudorapdity, identified-particle mean transverse momentum, identified-particle and charged-particle elliptic flow, and Hanbury-Twiss-Brown radii in both LHC 2.76 TeV  \cite{Alqahtani:2017jwl,Alqahtani:2017tnq} and RHIC 200 GeV heavy-ion collisions \cite{Almaalol:2018gjh}.  The framework includes both shear and bulk viscosities in addition to an infinite number of transport coefficient.  

In the version of the anisotropic hydrodynamics code used, the shear viscosity to entropy density is assumed to be constant and all other transport coefficients are self-consistently determined within a quasiparticle model with the temperature-dependent quasiparticle mass extracted from lattice QCD results for the entropy density \cite{Borsanyi:2010cj}.  For details of the framework we refer the reader to Ref.~\cite{Alqahtani:2017jwl,Alqahtani:2017tnq,Almaalol:2018gjh} and Ref.~\cite{Alqahtani:2017mhy} which presents a comprehensive review of the approach. More importantly this approach allows one to extend the dissipative hydrodynamical evolution to early times due to an infinite-order resummation in the inverse Reynolds number \cite{Strickland:2017kux}.

We begin the dissipative hydrodynamical evolution at $\tau = 0.2$ fm/c.  We use optical Glauber model to construct the initial energy density profile in the transverse plane and take the initial central temperature to be $T_0 = 600$ MeV and the shear viscosity to entropy density ratio to be $\eta/s = 0.2$.  These values were determined in prior studies by making fits to ALICE identified-particle spectra obtained in 2.76 TeV Pb-Pb collisions~\cite{Alqahtani:2017jwl,Alqahtani:2017tnq}.  The effective temperature profiles in different centrality classes were obtained from the anisotropic hydrodynamics code and were exported to disk.  These exported profiles were then used to construct 4-D interpolating functions which provided the spatiotemporal evolution of the effective temperature in each centrality class.  For more details concerning the hydrodynamical framework, fits, and comparisons to 2.76 TeV experimental data, we refer the reader to Ref.~\cite{Alqahtani:2017tnq}.

%%%%%%%%%%%%%%%%%%%%%%%%%%%%%%%%%%%%%%%%%%%%%%%%%%%%%%%%%%%%%%%%%%%%%

\section{Decay width and survival probability of bottomonia}
\label{s:model_survival}

Let us now turn to the description of the production of bottomonia and their behavior in the QGP.  The bottomonium states are produced in initial hard scattering processes during the very earliest stages of the heavy-ion collision. The spatial distribution of the production points in the transverse plane is assumed to follow that of the number of binary collisions, $N_{\rm coll}(x,y)$.  We assume a power-law transverse momentum ($p_T$) distribution of the $\Upsilon$'s obtained from PYTHIA simulations for $p+p$ collisions, scaled by the mass number of the colliding nuclei. Note that this kind of scaling implicitly assumes that the bottomonia do not ``flow'' with the medium and any $v_2$ that we obtain in our model will be purely due to the anisotropic escape mechanism. The initial $\Upsilon$ distribution for $p+p$ collision is assumed to be given by \cite{Zhou:2014hwa}
\begin{equation}
\label{ppjpsi}
\frac{d^2\sigma^{pp}_\Upsilon}{p_T\,dp_T\,dY} =
\frac{4}{3\langle p_T^2\rangle_{pp}}\!\left(1+{\frac{p_T^2}{\langle p_T^2\rangle_{pp}}}\right)^{\!\!-3}{\frac{d\sigma^{pp}_\Upsilon}{dY}} \, ,
\end{equation}
with $Y$ being the longitudinal rapidity in momentum space. Here $\langle p_T^2\rangle_{pp}(Y)=20(1-Y^2/Y_{\max}^2)\;$(GeV/$c)^2$, where $Y_{\max}={\rm cosh}^{-1}(\sqrt{s_{NN}}/(2m_{\Upsilon}))$ is the most forward rapidity of the bottomonia. Eventually, the momentum rapidity density follows a Gaussian distribution:
\begin{equation}
\label{gaussian}
\frac{d\sigma^{pp}_\Upsilon}{dY} = 
  \frac{d\sigma^{pp}_\Upsilon}{dY}\bigg|_{Y=0} {\rm e}^{-Y^2/0.33Y_{\max}^2}.
\end{equation} 
For our calculations, we integrate over $Y$ and consider the resulting $p_T$ distribution.

The formation of each bound bottomonium state requires a finite formation time $\tau_{\rm form}$. The value $\tau^0_{\rm form}$ of the latter in the bottomonium rest frame is assumed to be proportional to the inverse of the vacuum binding energy for each state. For the $\Upsilon (1S)$, $\Upsilon (2S)$, $\Upsilon (3S)$, $\chi_{b} (1P)$ and $\chi_{b}(2P)$ states we use $\tau^0_{\rm form} = 0.2, 0.4, 0.6, 0.4, 0.6$ fm/$c$, respectively~\cite{Krouppa:2016jcl}. In the laboratory frame, relative to which a bottomonium state with mass $M$ has transverse momentum $p_T$, the formation time becomes $\tau_{\rm form} = E_T\tau^0_{\rm form} / M$ with $E_T = \sqrt{p_T^2+M^2}$ being the transverse energy. 

After they are formed, since the bound $b\bar{b}$ states are color-neutral, their elastic scatterings on the color charges in the QGP are fewer and, because of their large rest mass, they propagate quasi freely, following nearly straight-line trajectories. We use the simulation results from QaHydro framework \cite{Alqahtani:2015qja} to obtain the temperature of the medium along the bottomonium trajectory. With this temperature, we compute the thermal decay widths of the bottomonium states, adopting the recent state-of-the-art estimations of in-medium dissociation of different bound $b\bar{b}$ states~\cite{Strickland:2011aa}, which are here considered for the case of a locally momentum-isotropic medium for simplicity.  For a given bound state, the 3-D Schr\"odinger equation is solved numerically with a temperature-dependent complex heavy-quark potential \cite{Margotta:2011ta}. The in-medium breakup rates for each state are then computed from the imaginary part of the binding energy as a function of the temperature $T/T_c$. The real part of the binding energy, on the other hand tells us when the state becomes completely unbound. We then set the temperature scale to $T_c = 160$ MeV. Below $T_c$, we assume that plasma screening effects are rapidly diminished due to transition to the hadronic phase and the widths of the states becoming approximately equal to their vacuum widths which in comparison to the in-medium widths are essentially zero.

The potential used for the solution of the 3-D Schr\"odinger equation is based on a generalized Karsch-Mehr-Satz potential \cite{Karsch:1987pv} obtained from the internal energy.  It includes an imaginary part which emerges due to Landau damping of the exchanged gluons in the hard-thermal-loop framework~\cite{Laine:2006ns,Dumitru:2007hy,Strickland:2011aa} but does not include effects of singlet to octet transitions \cite{Brambilla:2016wgg} and is explicitly based on the high-temperature limit of quantum chromodynamics.  By making this assumption we do not include any explicitly non-perturbative contributions to the imaginary part of the in-medium quarkonium potential. In general, the imaginary part of the potential encodes the break-up rate of heavy-quarkonium bound states and can be properly understood in the context of open quantum systems in which the heavy-quark system is quantum mechanically coupled to a thermal heat bath allowing for states to transition from bound states to the environmental sector of the in-medium density matrix~\cite{Akamatsu:2011se, Akamatsu:2012vt, Akamatsu:2014qsa, Katz:2015qja, Brambilla:2016wgg, Kajimoto:2017rel, Brambilla:2017zei, Blaizot:2017ypk, Blaizot:2018oev, Yao:2018nmy}. 

For a bottomonium with transverse momentum $p_T$ along the azimuthal $\phi_p$ direction, which is created with transverse co-ordinates $(x,y)$ at the formation time $\tau_{\rm form}$, the position at any future time $\tau$ is given by
\begin{equation}\label{future_position}
x' = x + v_T\tau'\cos\phi_p\quad, \quad y' = y + v_T\tau'\sin\phi_p,
\end{equation}
where $v_T=p_T/E_T$ is the bottomonium transverse velocity and $\tau' = \tau - \tau_{\rm form}$. The thermal decay width $\Gamma(T(x',y',\tau))$ is then modeled as 
%obtained from the solution to the Schr\"odinger equation, 
\begin{equation}
\Gamma(T(x',y',\tau)) = 
\left\{
\begin{array}{ll}
2 \Im  & \qquad {\rm for} \quad \Re >0, \\
10\;{\rm GeV}  & \qquad {\rm for} \quad \Re \le 0, \\
\end{array}
\right.
\end{equation}
where $\Im\equiv{\rm Im}[E_{\rm bind}(T(x',y',\tau'))]$ and $\Re\equiv{\rm Re}[E_{\rm bind}(T(x',y',\tau'))]$ denote the imaginary and real parts of the in medium binding energy, respectively, at a transverse position $(x',y')$ at time $\tau$.
The value of 10 GeV in the second case is chosen to rapidly dissociate states which are fully unbound and, in practice, the results are not significantly sensitive on this value as long as it is large enough to quickly melt the state under consideration.
The in-medium decay width for a given state, so obtained, thus determines its survival probability as it propagates in the medium. The final transmittance for a $b\bar{b}$ bound state labelled by $j$ is given by~\cite{Strickland:2011mw,Strickland:2011aa,Krouppa:2015yoa, Krouppa:2016jcl}
\begin{align}\label{transmittance_final}
{\cal T}_j(x,y,p_T,\phi_p) = \exp\bigg[\!&-\!\Theta(\tau_f - \tau^{\rm form}_j)\ \times \cr
&\hspace{-5mm}\int_{{\rm max}(\tau^{\rm form}_j,\tau_i)}^{\tau_f}\! d\tau'\;\Gamma_j \bigg(T(x', y';\,\tau')\!\bigg)\bigg], \cr & 
\end{align}
where $\Theta$ is the usual step function. The final time $\tau_f$ in the above equation is self consistently determined in the simulation as the proper time when the local effective temperature of the medium becomes less than the freeze out temperature $T_f = 130$ MeV.

From the above equation, we can obtain the transmitted spectra as a function of transverse momentum and azimuthal direction of all the produced bottomonium states,
\begin{equation}\label{spectra_pt_phi}
\dfrac{dN_j}{\,p_T\,dp_T\,d\phi_p}=\int dx\,dy\,n_{coll}(x,y)\frac{d^2\sigma^{pp}_\Upsilon}{d^2p_T\,}{\cal T}_j(x,y,p_T,\phi_p)
\end{equation}

%Note that the above form~\eqref{transmittance_final} of the escape probability of the produced bottomonium could alternatively be written in the form of the Beer--Lambert law used in Ref.~\cite{Jaiswal:2017dxp}, ${\cal T}=\exp(-\!\int\! n\,\sigma\,d\ell)$. The propagation time $\tau'$ clearly maps one-to-one to the in-medium path length $\ell=v_T\tau'$. In turn, the product of number density $n$ of the system, dissociation cross section $\sigma$, and velocity $v_T$ precisely give the rate of destruction $\Gamma$ of the bottomonium state under consideration.

%%%%%%%%%%%%%%%%%%%%%%%%%%%%%%%%%%%%%%%%%%%%%%%%%%%%%%%%%%%%%%%%%%%%%

\section{Results and discussions}
\label{s:results}

In this section, we numerically evaluate Eq.~(\ref{spectra_pt_phi}) to calculate the elliptic flow of bottomonia due to escape probability through a medium having an anisotropic shape. We consider Pb$\,+\,$Pb collisions at $\sqrt{s_{\rm NN}}=2.76$~TeV in $40-50$\% centrality class which corresponds to an average impact parameter of $10.41$~fm~\cite{Abelev:2013qoq}. 

%The energy density in the transverse plane is generated from a Glauber model with Woods-Saxon nuclear density distribution, following a linear combination of the spatial profiles of the number of participant nucleons $N_{\rm part}$ and number of binary collisions $N_{\rm coll}$ as $\varepsilon_i(x,y) \propto 0.85N_{\rm part}(x,y)+0.15N_{\rm coll}(x,y)$~\cite{Strickland:2011aa}. The inelastic nucleon-nucleon interaction cross section is taken to be $\sigma_{NN}=62$~mb and the initial energy density in the central cell of central Pb+Pb collisions at $\tau_i=0.4$~fm is taken to be 85~GeV/fm$^3$, which corresponds to an initial temperature of 480 MeV.

%%%%%%%%%%%%%%%%%%%%%%%%%%%%%
\begin{figure}[t]
\begin{center}
\includegraphics[width=\linewidth]{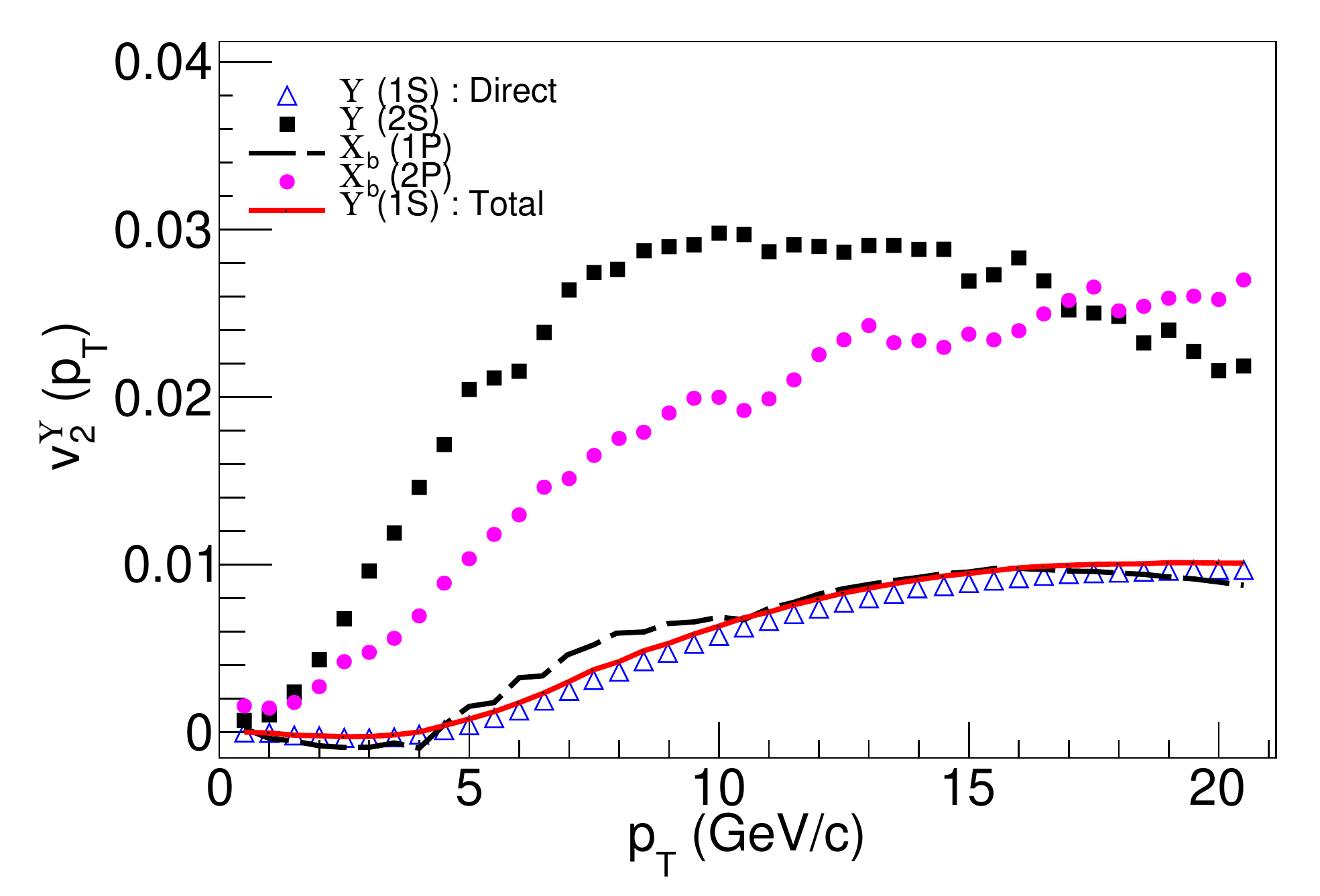}
\end{center}
\vspace{-0.8cm}
\caption{Transverse momentum dependence of elliptic flow parameter for different bottomonium states for Pb$\,+\,$Pb collisions at $\sqrt{s_{\rm NN}}=2.76$~TeV in $40-50$\% centrality class. For $\Upsilon (1S)$, directly produced states and the inclusive yield including feed down contributions are shown separately.} 
\label{v2_comp}
\end{figure}
%%%%%%%%%%%%%%%%%%%%%%%%%%%%%

In order to compare to the experimental results, we integrate over the entire temperature profile in the transverse plane, to obtain the weighted average ``raw'' spectra for each bottomonium state, as shown in Eq.~\eqref{spectra_pt_phi}. To account for the post-QGP feed down of the excited states, we use a $p_{T}$-averaged feed down fraction obtained from a recent compilation of $p+p$ data at LHC. The inclusive spectra for $\Upsilon (1S)$  is then calculated from a linear superposition of the raw spectra for each state: 
\begin{equation}\label{in_spec}
{dN^{all}_{\Upsilon(1S)}\over {p_T dp_{T}d\phi_p}}=\sum_{j} f_{j}{dN_{j}\over {p_Tdp_{T}d\phi_p}},
\end{equation}
where $j$ is used to label the different bound states of bottomonia. The contribution from different states are as follows: $f_{2S\rightarrow 1S}=8.6\%$, $f_{3S\rightarrow 1S}=1\%$, $f_{1P\rightarrow 1S}=17\%$, $f_{2P \rightarrow 1S}=5.1\%$ and $f_{3P \rightarrow 1S}=1.5\%$ as adopted from \cite{Krouppa:2016jcl}. Since contribution from $3S$ and $3P$ states are small, we include their percentage contributions in $2S$ and $2P$ states, respectively. The inclusive spectra so constructed is then used to calculate the elliptic flow $v_2$ of $\Upsilon(1S)$. 
Note that, while considering feed down, the transverse momentum of the mother and daughter bottomonium states are assumed to be the same. This assumption can be justified by considering the average $p_T$ value of the mother excited states and the daughter $1S$ states. Due to the large mass of the bottomonium states, we find that the mean $p_T$ value of the mother excited states and the daughter $1S$ states are almost identical.

%From \eqref{transmittance_final}, we can obtain the transmittance probability as a function of transverse momentum and azimuthal direction of all the produced bottomonium states,
%\begin{equation}\label{trans_pt_phi}
%{\cal T}_j(p_T,\phi_p) = \int dx\,dy\,{\cal T}_j(x,y,p_T,\phi_p).
%\end{equation}

Our main objective is to find to $v_{2}(p_{T})$ of $\Upsilon(1S)$ states generated from anisotropic dissociation in the plasma. However no such experimental measurement is available for the same, with which our model results can be contrasted. Hence to check the viability of our model, in addition to $v_{2}(p_{T})$ we also calculate the $p_T$ dependence of nuclear modification factor, $R_{AA}(p_{T})$ that has been widely studied at LHC. 

%The transverse momentum dependent nuclear modification factor of $j$-th bound state
%\begin{equation}\label{RAA}
%R_{\rm AA}^j(p_T) = \int d\phi_p\,{\cal T}_j(p_T,\phi_p).
%\end{equation}
%Similarly, using Eq.~\eqref{trans_pt_phi} we obtain the elliptic flow $v_2$ of $j$-th bound state
%\begin{equation}\label{v2_trans}
%v_n^j(p_T) \equiv 
%\dfrac{\displaystyle{\int_{-\pi}^{\pi}}d\phi_p\,\cos(2\phi_p)\,{\cal T}_j(p_T,\phi_p)\,\dfrac{dN_j}{dy\,p_T\,dp_T\,d\phi_p}}
%{\displaystyle{\int_{-\pi}^{\pi}}d\phi_p\,{\cal T}_j(p_T,\phi_p)\,\dfrac{dN_j}{dy\,p_T\,dp_T\,d\phi_p}}.
%\end{equation}
Using Eq.~\eqref{spectra_pt_phi} and following standard formulae, we thus calculate $R_{\rm AA}^j(p_T)$ and $v_2^j(p_T)$ of $j$-th bound state. For inclusive $\Upsilon(1S)$ states we make use of Eq.~\eqref{in_spec}. While calculating $v_2$, we have not accounted for the reaction plane angle as we have considered smooth optical Glauber model initial conditions.

%%%%%%%%%%%%%%%%%%%%%%%%%%%%%
\begin{figure}[t]
\begin{center}
\includegraphics[width=\linewidth]{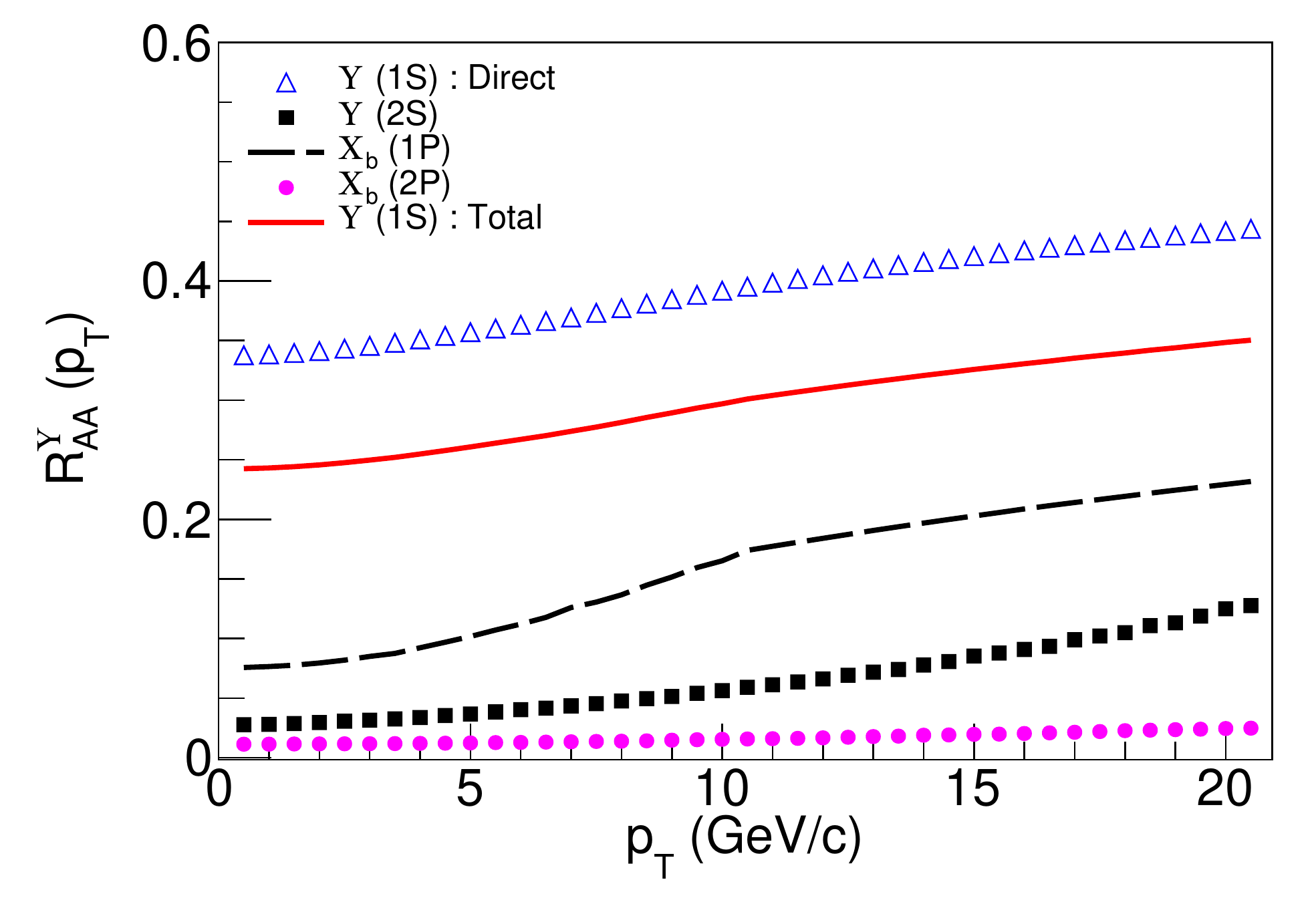}
\end{center}
\vspace{-0.8cm}
\caption{Transverse momentum dependence of nuclear modification factor $R_{AA}(p_{T})$ for different bottomonium states for Pb$\,+\,$Pb collisions at $\sqrt{s_{\rm NN}}=2.76$~TeV in $40-50$\% centrality class. For $\Upsilon (1S)$, directly produced states and the inclusive yield including feed down contributions are shown separately} 
\label{RAA_comp}
\end{figure}
%%%%%%%%%%%%%%%%%%%%%%%%%%%%%

The $p_T$ dependence of $v_2$ for different directly produced bottomonium states for Pb$\,+\,$Pb collisions at $\sqrt{s_{\rm NN}}=2.76$~TeV in $40-50$\% centrality class is shown in Fig.~\ref{v2_comp}. For each state, $v_2$ initially increases with $p_T$ and gradually tends to saturate. A reverse hydrodynamic mass ordering between the different bottomonium states is observed with more massive state having larger $v_2$. At a given $p_T$, owing to weaker binding and hence broader decay width, $v_2$ is larger for the excited states, as was found in Ref.~\cite{Du:2017qkv}. However this ordering according to the binding energies gets broken for the $\chi_{b}(2P)$ state, which is seen to acquire smaller $v_2$ than $\Upsilon$(2S) state. Even though $\chi_{b}(2P)$ remains practically unbound during the entire evolution of the plasma, due to large intrinsic formation time, the suppression effects remain operative for a shorter period, resulting in lesser flow. The non-monotonic nature of $v_{2}(p_{T})$ for the excited states can be attributed to the melting of those states beyond which the decay width is abruptly set to 10 GeV. Note that such non-monotonocity is not seen for $\Upsilon(1S)$ states, for which the melting temperature is around 900 MeV.\footnote{The present results on $v_2(p_T)$ of different bottomonium states, with realistic evolution dynamcis are somewhat different compared to our previous estimations (arXiv:1809.06235v1 [hep-ph]), where we used a parametrized transverse exapnsion profile.} 

In Fig.~\ref{v2_comp}, we also show the inclusive $v_2$ of the $1S$ state which takes into account feed-down. As the largest contribution to inclusive $v_2$ of $1S$ states comes from the decay of $1P$ states, it is close to the direct $v_2$ of $1S$ states. At very high $p_T \gtrsim 10$ GeV, $v_2$ tends to decrease with increasing $p_T$, an effect which is more prominent for excited states. This reduction can be attributed to the competition between the dynamics of the plasma and that of the bottomonium state under consideration. Due to both its velocity as well as its dilated formation time in the plasma frame, a bottomonium with large $p_T$ escapes faster from the plasma, spending less time inside and therefore the suppression effect is less important, leading to a reduced $v_2$.

% At a given $p_T$, owing to weaker binding and hence broader decay width, $v_2$ is larger for the excited states, as was found in Ref.~\cite{Du:2017qkv}. 
%As nearly $30 \%$ of the measured $\Upsilon (1S)$ come from the decay of excited states, inclusive $v_2$ of $1S$ is reasonably larger than that of  directly produced $1S$ states. At very high $p_T \gtrsim 10$ GeV, $v_2$ tends to decrease with increasing $p_T$, an effect which is more prominent for excited states. This reduction can be attributed to the competition between the dynamics of the plasma and that of the bottomonium state under consideration. Due to both its velocity as well as its dilated formation time in the plasma frame, a bottomonium with large $p_T$ escapes faster from the plasma, spending less time inside and therefore the suppression effect is less important, leading to a reduced $v_2$.

%It is important to test the sensitivity of our model results to various input parameters.

Before moving forward, it might be interesting to look at the $p_T$ dependence of $R_{AA}$ within the same model framework. In Fig.~\ref{RAA_comp}, we show $R_{AA}(p_T)$ for different bottomonium states with same set of input parameters. For any given state, as expected $R_{AA}$ is seen to increase with $p_T$ due to a faster escape from the dense medium. The suppression pattern for different states exhibit a monotonic trend. Owing to weakest binding the $2P$ states are seen to have largest suppression. For inclusive $1S$ states the results are in line with the previous calculations~\cite{Krouppa:2017jlg}. 

%%%%%%%%%%%%%%%%%%%%%%%%%%%%%%%%%%%
 \begin{figure}[t]
\begin{center}
\includegraphics[width=\linewidth]{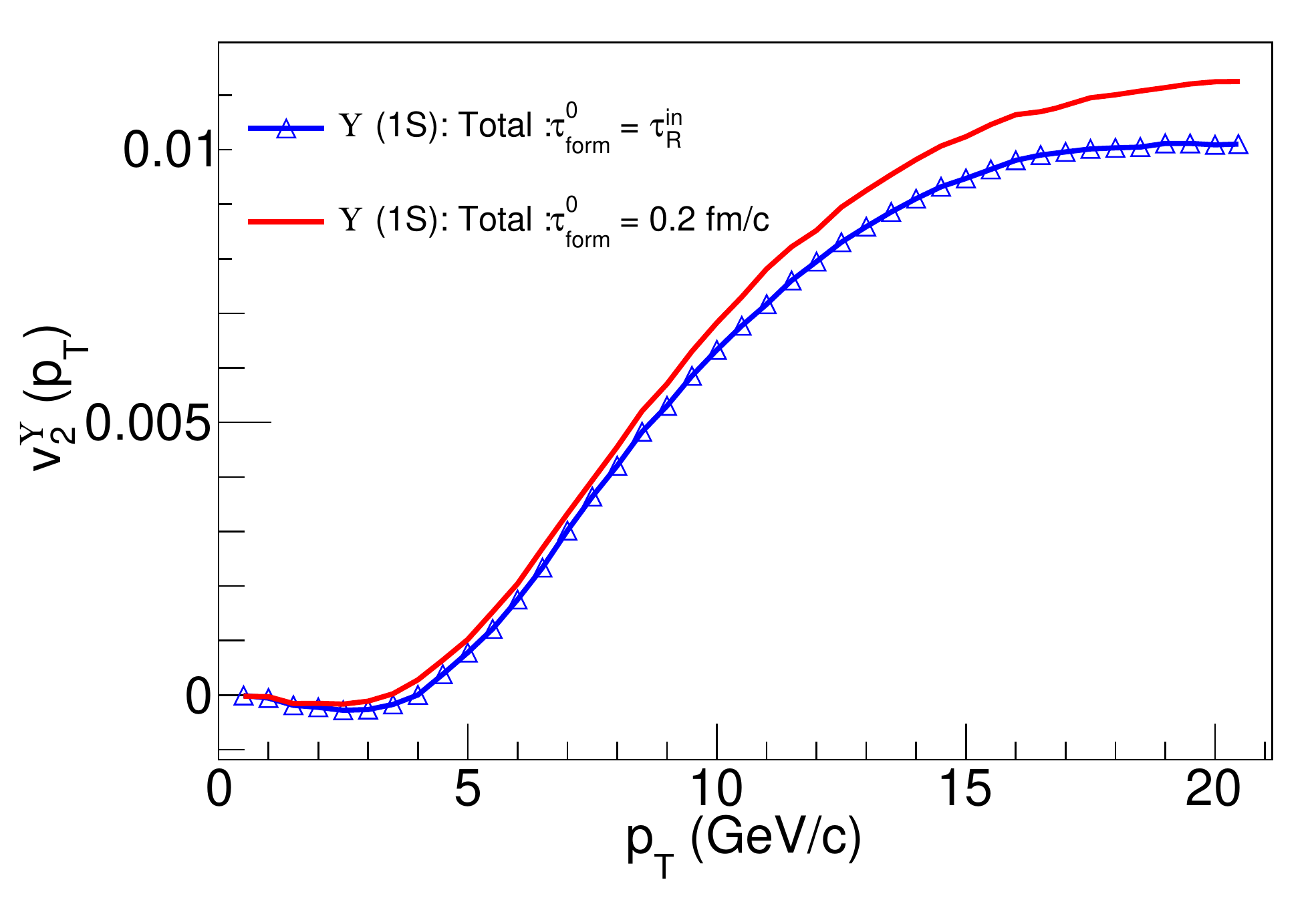}
\end{center}
\vspace{-0.8cm}
\caption{Transverse momentum dependence of $v_2$ of $\Upsilon (1S)$ including feed down contributions from higher excited states for Pb$\,+\,$Pb collisions at $\sqrt{s_{\rm NN}}=2.76$~TeV.The two curves correspond to two different assumptions for the intrinsic formation time of the excited states (see text for details).} 
\label{v2_formation}
\end{figure}
%%%%%%%%%%%%%%%%%%%%%%%%%%%%%%%%%%%%%%%%%%%%%%%%%%%%%%%%

%In Fig.~\ref{v2_comp_beta}, we show $v_2(p_T)$ for different combinations of the input parameters $\beta_0^{\rm BW}$ and $\beta_2^{\rm BW}$ controlling the medium expansion for Pb$\,+\,$Pb collisions at $\sqrt{s_{\rm NN}}=2.76$~TeV in $40-50$\% centrality class. Instead of considering all bottomonium states, we only report the directly produced $1S$ states for this purpose. We find that a rather stronger $v_2$ results as soon as transverse expansion of the medium is included, i.e.\ for $\beta_0^{\rm BW}\neq 0$. However in presence of transverse expansion, $v_2$ is much less sensitive to the parameter $\beta_2^{\rm BW}$. Increasing $\beta_2^{\rm BW}$ leads to a decrease of $v_2$. This may be understood from the fact that, since the medium flow does not impart any $v_2$ to bottomonium, increasing $\beta_2^{\rm BW}$ leads to a faster decrease of the spatial anisotropy of the medium and hence to a diminution of $v_2$ generated from anisotropic escape.

It might be interesting to check the sensitivity of the results to different model inputs. Instead of showing different bottomonium states separately, we only report the inclusive $1S$ states for this purpose. To check the effect of intrinsic resonance formation time on resulting flow, we calculate $v_{2}(p_{T})$ assuming similar intrinsic formation time, $\tau^{0}_{form} = 0.2$ fm/$c$, for all states. The result is contrasted with our previous calculation (different intrinsic formation times) in Fig.~\ref{v2_formation}. If all the bottomonium states are assumed to be formed at same time, they experience screening effects for a longer time, resulting in larger $v_{2}$, an effect more prominent at higher $p_{T}$. 

%%%%%%%%%%%%%%%%%%%%%%%%%%%%%
\begin{figure}[t]
\begin{center}
\includegraphics[width=\linewidth]{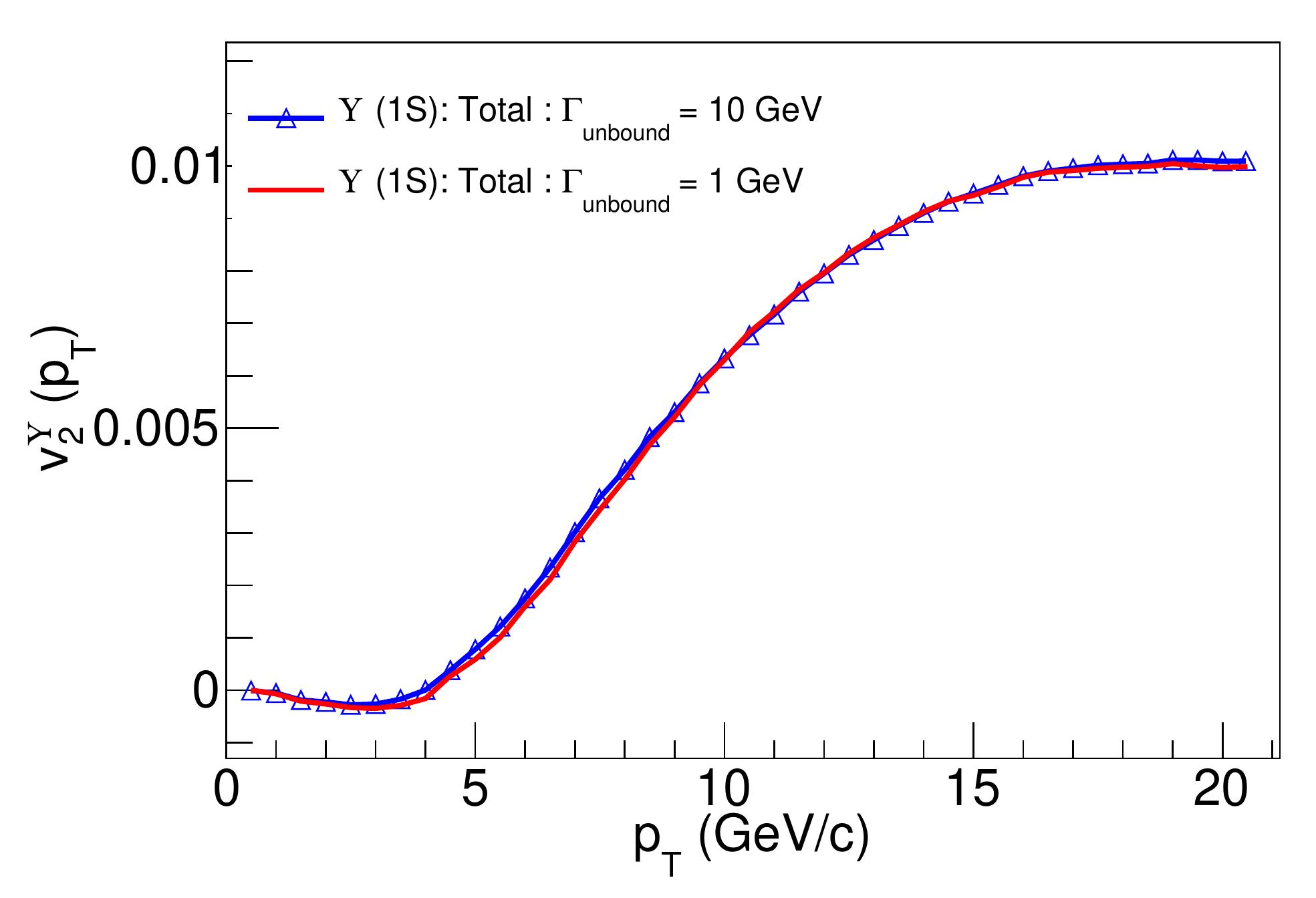}
\end{center}
\vspace{-0.8cm}
\caption{Transverse momentum dependence of $v_2$ of $\Upsilon (1S)$ including feed down contributions from higher excited states for Pb$\,+\,$Pb collisions at $\sqrt{s_{\rm NN}}=2.76$~TeV, for two different values of the decay width of unbound states.} 
\label{v2_unbound}
\end{figure}
%%%%%%%%%%%%%%%%%%%%%%%%%%%%%

Next we check the effect of decay widths of the unbound states. In Fig.~\ref{v2_unbound}, we show the inclusive $v_2$ of $1S$ states for two arbitrary values of $\Gamma_{\infty}$ 1~GeV and 10~GeV. Since $70\%$ of the measured $\Upsilon (1S)$ states are direcly produced which are always bound inside fireball, the resulting $v_2$ values are practically same.

\begin{figure}[t]
\begin{center}
\includegraphics[width=\linewidth]{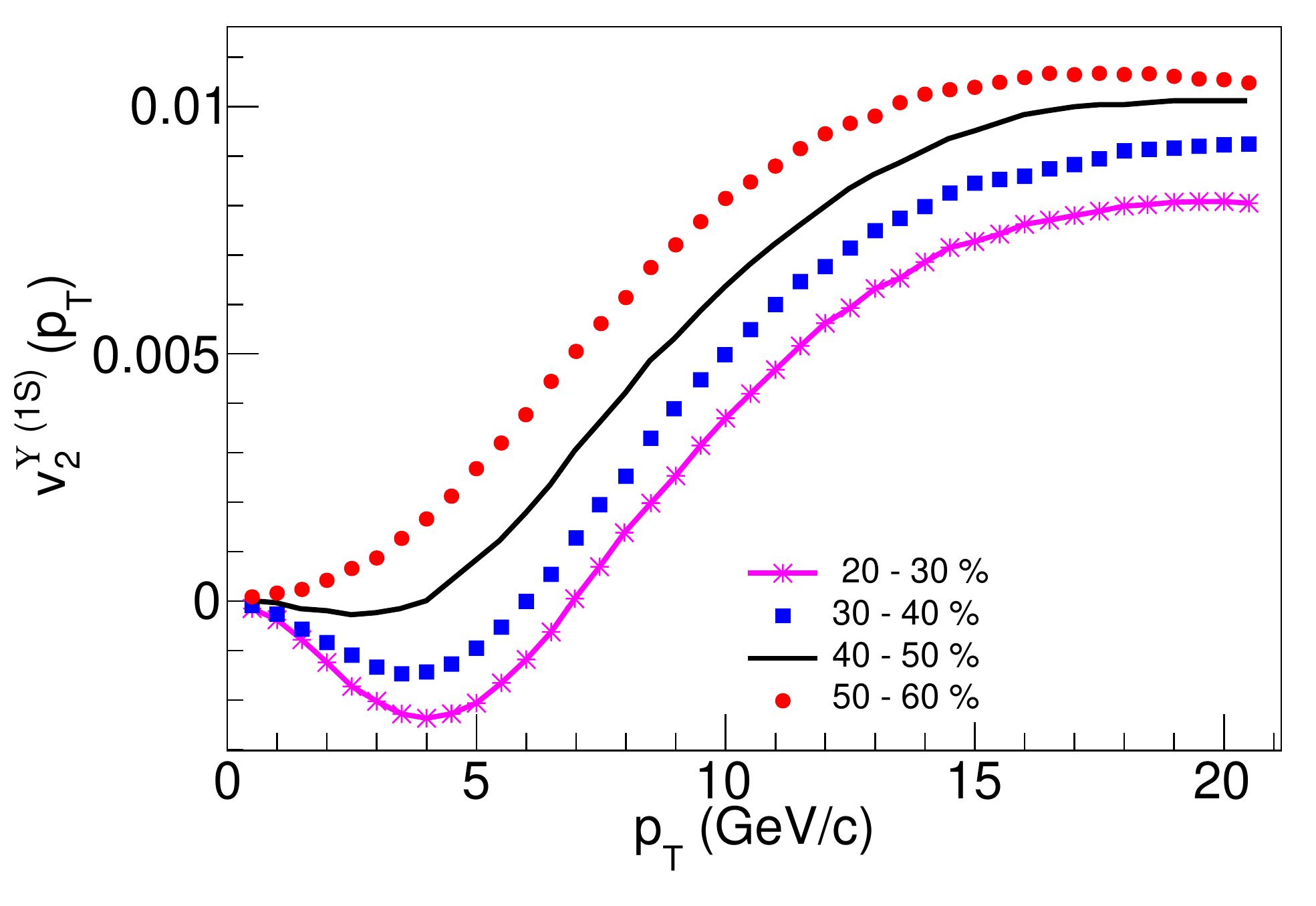}
\end{center}
\vspace{-0.8cm}
\caption{Transverse momentum dependence of $v_2$ of $\Upsilon (1S)$ including feed down contributions from higher excited states for different centrality for Pb$\,+\,$Pb collisions at $\sqrt{s_{\rm NN}}=2.76$~TeV.} 
\label{v2_comp_cent}
\end{figure}

Finally, in Fig.~\ref{v2_comp_cent}, we show the transverse momentum dependence of $v_2$ of $\Upsilon (1S)$ including feed down contributions from higher excited states for different centrality for Pb$\,+\,$Pb collisions at $\sqrt{s_{\rm NN}}=2.76$~TeV. We see that there is a non-monotonic behaviour of $v_2$ with respect to centrality and maximum $v_2$ is obtained for $50 - 60\%$ centrality. While eccentricity increases with centrality, the breakup rates decreases with temperature and goes to zero if the bottomonium has escaped. As we increase the impact parameter, the central temperature decreases and the resulting bottomonium momentum anisotropies decrease because of the decrese in thermal decay width $\Gamma(T)$. However spatial anisotropy increases at a much faster rate than the decreasing of thermal decay width leading to the monotonic behaviour of bottomonium $v_2$ with respect to centrality as seen in Fig.~\ref{v2_comp_cent}.

%%%%%%%%%%%%%%%%%%%%%%%%%%%%%
%%%%%%%%%%%%%%%%%%%%%%%%%%%%%

%We see that there is a non-monotonic behaviour of $v_2$ with respect to centrality and maximum $v_2$ is obtained for $30-40\%$ centrality. This may be attribute to the fact that while eccentricity increases with centrality, the breakup rates decreases with temperature and goes to zero if the bottomonium has escaped. As we increase the impact parameter, the central temperature decreases and the resulting bottomonium momentum anisotropies decrease because of the decrese in thermal decay width $\Gamma$. The competition between increasing spatial anisotropy and decreasing thermal decay width leads to the non-monotonic behaviour of bottomonium $v_2$ with respect to centrality as seen in Fig.~\ref{v2_comp_cent}.

At this point, it is worth mentioning that for bottomonia that are in motion relative to the expanding quark-gluon plasma, the in-medium dissociation depends on the effective local temperature that the quarkonia experiences due to the relativistic Doppler effect \cite{Escobedo:2013tca, Hoelck:2016tqf}. As a consequence, bottomonia with a higher velocity relative to the QGP are expected to be more affected. For effective hydrodynamic expansion, that we consider in the present case, the QGP is initially at rest so the high-$p_T$ bottomonia would indeed be more suppressed. However, these high-$p_T$ bottomonia have a larger formation time and hence the bottomonium experiences smaller temperatures due to rapid cooling of the medium. Since the dissociation width is smaller at lower temperatures, the suppression is lesser leading to smaller $v_2$. Moreover, the averaging over redshifted and blueshifted regions can potentially wash out the influence of Doppler corrected temperature and therefore we expect this effect on $v_2$ to be small. We postpone this analysis for future work.

Before closing we note that the ALICE Collboration has very recently reported the first measurement of $v_2$ of inclusive $\Upsilon(1S)$ states in Pb$\,+\,$Pb collisions at $\sqrt{s_{\rm NN}} = 5.02$ TeV, at forward rapidity~\cite{Acharya:2019hlv}. Due to paucity of statistics $v_2(p_T)$ has been calculated for a large centrality interval $5 - 60 \%$. The measured $v_2$ values is consistent with zero and with small positive values predicted by the transport models within large uncertainty. Of course we can not make a one-to-one correspondence with our present calculations, due to difference in beam energy and rapidity interval.

%%%%%%%%%%%%%%%%%%%%%%%%%%%%%%%%%%%%%%%%%%%%%%%%%%%%%%%%%%%%%%%%%%%%%

\section{Summary and conclusion}

In this paper, we have provided a quantitative prediction for the elliptic flow of bottomonia produced in mid-central collisions in $\sqrt{s_{\rm NN}}=2.76$~TeV Pb$\,+\,$Pb collisions at LHC via an anisotropic escape mechanism. We employed the Glauber model to generate initial distribution of energy density in the plane transverse to the beam axis. Using temperature-dependent decay widths for bottomonium states, we calculated their survival probability when traversing through the hot and dense anisotropic medium formed in non-central collisions. For the expansion of the fireball, we have used results from the recently developed 3+1d quasiparticle anisotropic hydrodynamic simulation. We also accounted for the feed down contribution to the bottomonium ground state from higher excited states. We found that the transverse momentum dependence of the elliptic flow of bottomonia is of the level of few percent, consistent with the finding of Ref.~\cite{Du:2017qkv}. We also found a monotonic behaviour of $v_2$ with respect to centrality. For completeness, we have also calculated the $p_T$ dependence of $R_{AA}$ which is consistent with the existing theoretical and experimental estimates.

Looking forward, it will be interesting to consider the effect of medium-induced transitions between bound states which is predicted from the open quantum system approach \cite{Borghini:2011yq, Borghini:2011ms, Akamatsu:2011se, Dutta:2012nw, Brambilla:2016wgg, Blaizot:2017ypk}. In this ``state reshuffling'' scenario, transitions between various bound states become possible which counteracts the usual suppression picture by allowing for the re-formation of otherwise suppressed states even above the hadronization temperature. Since the excited states of bottomonium acquire more elliptic flow due to anisotropic escape mechanism, one may expect to generate larger flow owing to the feed-down from excited states. We leave these questions for future work.

%%%%%%%%%%%%%%%%%%%%%%%%%%%%%%%%%%%%%%%%%%%%%%%%%%%%%%%%%%%%%%%%%%%%%

\begin{acknowledgments}

N.\,B.\ acknowledges support by the Deutsche For\-schungs\-gemeinschaft (DFG) through the grant CRC-TR 211 ``Strong-interaction matter under extreme conditions''. A.\,J.\  is supported in part by the DST-INSPIRE faculty award under Grant No. DST/INSPIRE/04/2017/000038.  M.\,S. is supported by the U.S. Department of Energy, Office of Science, Office of Nuclear Physics under Award No.~DE-SC0013470. This research was supported by the Munich Institute for Astro- and Particle Physics (MIAPP) of the DFG cluster of excellence ``Origin and Structure of the Universe''. P. P. B. would like to thank W. Sheikh and T. K. Nayak for an enriching discussion of latest ALICE measurements.  

\end{acknowledgments}

%---------------------------------------------------------------------

\end{document}